# Ligand-Controlled Phonon Dynamics in CsPbBr$_3$ Nanocrystals Revealed by Machine-Learned Interatomic Potentials

Seungjun Cha,[1] Chen Wang,[2,3,*] Victor Fung,[4,*] Guoxiang Hu[1,5,*]

[1] School of Materials Science and Engineering, Georgia Institute of Technology, Atlanta, GA 30332, USA

[2] Department of Chemistry and Biochemistry, Queens College, City University of New York, New York, NY 11367, USA

[3] The Graduate Center, City University of New York, New York, NY 10016, USA

[4] School of Computational Science and Engineering, Georgia Institute of Technology, Atlanta, GA 30332, USA

[5] School of Chemistry and Biochemistry, Georgia Institute of Technology, Atlanta, GA 30332, USA

*Corresponding author.

Email: chen.wang@qc.cuny.edu, victorfung@gatech.edu, emma.hu@mse.gatech.edu

## ABSTRACT

Halide perovskite nanocrystals are leading candidates for next-generation optoelectronics, yet the role of surface ligands in controlling their phonon dynamics remains poorly understood. These dynamics critically govern nonradiative relaxation, energy up-conversion, and phonon-assisted anti-Stokes emission. Conventional *ab initio* methods, while accurate, are computationally infeasible for experimentally relevant nanocrystal sizes that require thousands of atoms to capture realistic ligand shells and dynamic disorder at finite temperatures. Here, we introduce a machine-learned interatomic potential fine-tuned on small CsPbBr$_3$ nanocrystals with diverse ligands, enabling accurate prediction of ligand-induced phonon properties far beyond the spatial and temporal scales of *ab initio* methods. We find that both cationic and anionic ligands systematically redshift Pb-Br-Pb stretching modes while blueshifting the PbBr$_6^{4-}$ octahedral rotation mode, with stronger overall effects for anionic passivation. Notably, anionic ligands stiffen the rotation mode non-monotonically with respect to the ligand binding energy. Our findings reveal important roles of cationic and anionic ligands in modulating key dynamic modes of halide perovskite nanocrystals associated with detrimental nonradiative losses, offering mechanistic insights and design principles for high-performance perovskite nanocrystal optoelectronics.

## INTRODUCTION

Halide perovskite nanocrystals (NCs) have emerged as the next-generation optoelectronic materials due to narrow emission linewidth, high defect tolerance, and low-cost solution-processability, which enables precise control over compositions and sizes for property tuning[1,2]. In particular, colloidal NCs with sizes in the quantum confinement regime largely benefit from enhanced exciton binding energies and carrier localization[3], making them attractive candidates for light-emitting applications such as LEDs[4], lasers[5], and single-photon sources[6]. Colloidal synthesis of halide perovskite NCs typically employs organic ligands, which cap the surface of the inorganic core to provide colloidal stability in solvents. These ligands play a pivotal role in determining the optoelectronic performance of the NCs. Surface ligands are essential for passivating trap states in the bandgap (i.e., deep traps), which promote nonradiative recombination pathways that harm photoluminescence quantum yield (PLQY). Deep traps predominantly arise from surface defects[7,8] and become especially detrimental to NCs with a high surface-to-volume ratio[9], hindering near-unity PLQY ideal for light-emitters.

More recently, the role of surface ligands in controlling phonon dynamics has been increasingly recognized. For example, conjugated molecular multipods have been shown to suppress dynamic disorder in FAPbBr$_3$ by stiffening the lattice through hydrogen bonding and van der Waals interactions with the surface FA molecules, observed as a blueshift of the low-energy PbBr$_6^{4-}$ rotation mode[10]. Low-energy phonon modes in CsPbBr$_3$ such as rotation, bending and stretching modes of Pb and Br, which directly couple to surface ligands, have been identified to govern processes such as energy up-conversion[11], phonon-assisted anti-Stokes emission[12], and carrier cooling[13]. These processes have direct implications for radiative efficiency, performance, and applications of perovskite nanocrystal optoelectronics. Thus, fundamental understanding of nanocrystal surfaces, particularly how ligands affect low-energy phonons, emerges as a key challenge for designing these unique materials systems. Yet, despite their importance, the theoretical understanding of ligand-controlled phonon dynamics remains limited, in part due to the lack of computationally tractable methods.

Conventional *ab initio* studies model halide perovskite NCs as periodic slabs[7,14–16], in which a large vacuum is added along one axis while maintaining 2D periodicity along the others. While computationally efficient, slab models are inherently unsuitable for capturing the complexity of NCs, which are quantum-confined in all three dimensions. As a result, they fail to capture key properties including size effect, site-dependent properties (at corners and edges), and interactions between opposing surfaces present in a finite system. By contrast, realistic NC models naturally account for these effects but come with a prohibitive computational cost: due to the lack of periodicity, the number of atoms scales exponentially with NC size, making *ab initio* methods such as density functional theory (DFT) impractical beyond modest spatial and temporal scales. The incorporation of surface ligands exacerbates the problem. For example, a realistic 4 nm CsPbBr$_3$ NC model capped with 50% surface coverage of methylammonium and benzoate has ~5,000 atoms, which is far beyond the tractable limit for conventional DFT. The compute-heavy nature of phonon calculations and molecular dynamics forces this limit to be even lower. Consequently, phonon studies using NC models have largely been restricted to ligand-free systems[13] or those with limited coverage[17]. This calls for a more computationally efficient alternative to fully capture the effect of surface ligands in realistic NCs.

To overcome these limitations, we turn to universal machine-learned interatomic potentials (MLIPs) to achieve near-DFT accuracy at a fraction of computational cost. While many existing MLIPs are largely pretrained on bulk crystals[18–22], we find that they can be effectively fine-tuned on small NC models with DFT-feasible sizes and representative ligands, which extrapolate well outside the training regime. Here, we systematically compute the phonon density of states of the representative NC models across different sizes to reveal ligand-controlled phonon dynamics in $CsPbBr_3$ NCs. We discuss the mode-selective effect of cationic and anionic ligands in modulating characteristic Pb-Br modes associated with nonradiative loss. We highlight the non-monotonic effect of anionic ligand binding energies on the stiffening of $PbBr_6^{4-}$ octahedral rotation mode, which was further studied using molecular dynamics simulations. These results provide insights into the design principles for tailoring surface chemistry to control lattice disorder in halide perovskite optoelectronics.

## RESULTS AND DISCUSSIONS

**Fine-tuning a universal machine-learned interatomic potential**

We first constructed a systematic set of $CsPbBr_3$ NC models with varying supercell sizes, surface passivation schemes, and ligand coverage (Figure 1). The three representative cases were (1) bare NCs, (2) methylammonium (MA)-capped NCs, and (3) more realistic mixed-ligand NCs capped with both MA and anionic ligands. (Table S1 provides a full list of ligands used in this study). Here,

MA represents popular ammonium-derivatives such as oleylammonium, which is almost always present on the surface of halide perovskite NCs during typical synthesis[15,23,24]. All NC models were constructed with Cs-Br termination and strict charge neutrality by removing excess Cs atoms on the surface. In ligand-capped models, MA and anionic ligands randomly replaced surface Cs and Br atoms, respectively, while maintaining specified ligand coverage. To capture an even wider phase space and local environments during training, atomic positions were randomly perturbed in proportion to their covalent radii. For mixed-ligand NCs, a short pre-relaxation step using the pretrained MLIP was applied in order to prevent steric crowding during subsequent geometry relaxation. The train and test datasets were then generated from the geometry relaxation trajectories of the NC models using Vienna Ab initio Simulation Package (VASP),[25] naturally yielding many near-equilibrium structures favorable for phonon calculations.

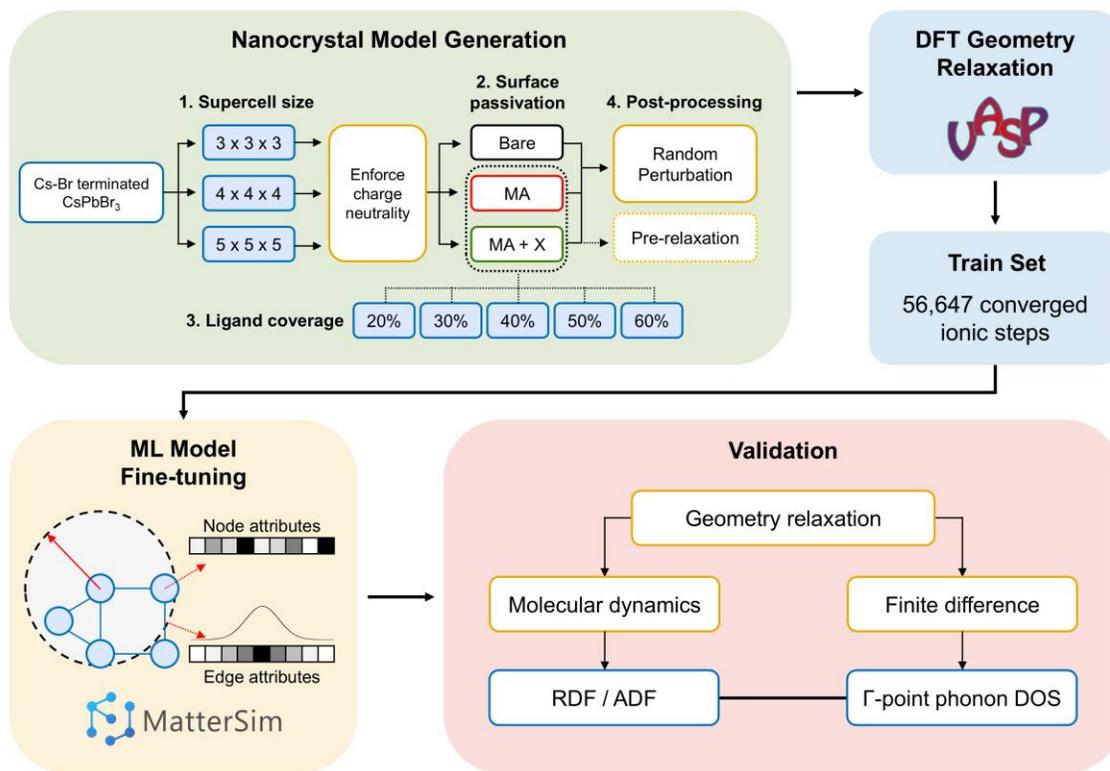

Figure 1. Overview of this work. NC models were generated by varying (1) supercell size, (2) surface passivation scheme, and (3) ligand coverage, while enforcing charge neutrality. The models were relaxed with the DFT level of theory using VASP, generating a training set of 56,647 converged ionic steps. MatterSim-v1.0.0-1M was fine-tuned via the MatterTune package [26] and validated against the test set using radial/angular distribution functions and Γ-point phonon density of states as validation metrics.

For this work, MatterSim-v1.0.0-1M[19] was selected for fine-tuning because conservative models were shown to have a good performance in phonon prediction[27]. We first assessed the predictive performance of the pretrained MatterSim on our train dataset as the baseline. For per atom energies, the pretrained model already achieves high accuracy with a near-unity $R^2$ and an RMSE of 0.0229 eV/atom (Figure 2a, left). However, its force predictions were highly unreliable with an R² of merely 0.3694. The culprit for such a low correlation was the low-force regime, as illustrated in the zoomed-in parity plot (Figure 2a, right). Since our dataset comes from geometry relaxation, the vast majority of the force datapoints are clustered near equilibrium. In this region, the pretrained model completely loses its predictive power, with noticeable deviations centered around zero that render it unsuitable for phonon calculations near equilibrium. This behavior is expected because universal potential was mainly trained on bulk crystals, where local environments can differ substantially from those at nanocrystal interfaces. Fine-tuning on our nanocrystal dataset was therefore not only necessary but also highly effective. Both energy and force predictions improved by an order of magnitude relative to the baseline model. In particular, force accuracy in the near-equilibrium region increased dramatically, achieving an $R^2$ of 0.9878 and RMSE of 0.0230 eV/Å (Figure 2b, middle and right). This demonstrates that local environment learning via message-passing embedded in many universal potentials is transferable to nanocrystal systems, and that fine-tuning substantially enhances force prediction fidelity for further downstream tasks.

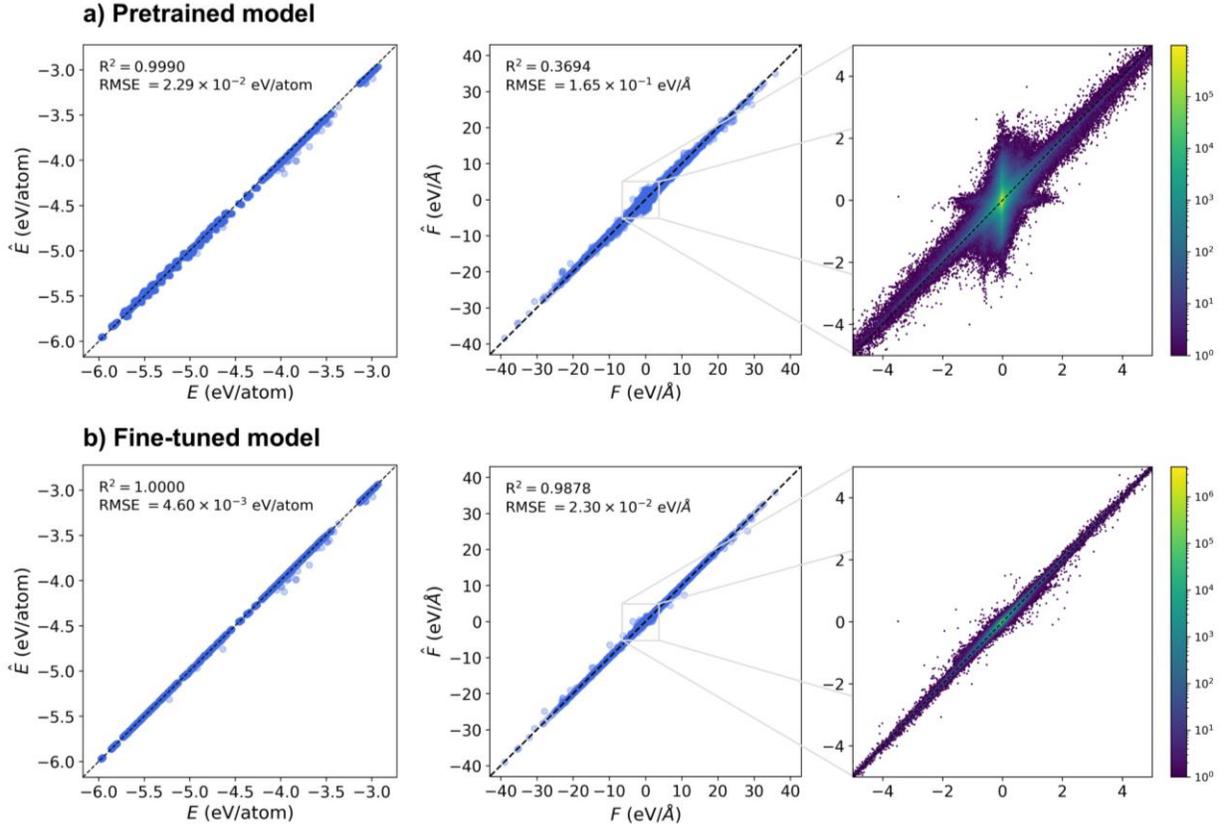

Figure 2. Test dataset parity plots of predicted vs. reference energies, $E$, and forces, $F$, for (a) the pretrained model and (b) the fine-tuned model. The left panels show energy predictions, the middle panels show force predictions, and the right panels present zoomed-in force parity plots with point density indicated by the color bar.

To further validate the fine-tuned MLIP, we compared its structural and vibrational predictions against DFT benchmarks (Figure 3). The Pb-Br radial distribution function (RDF) and Pb-Br-Pb angular distribution function (ADF), obtained from short NVT trajectories of 2×2×2 NC capped with 50% benzoate (BzO) and methylammonium (MA), show excellent agreement between the two methods (Figure 3a,b). Both methods have nearly identical positions and relative intensities of the major peaks characteristic to orthorhombic $CsPbBr_3$,[28–30] indicating that the MLIP can well reproduce the local coordination environments in $CsPbBr_3$ NCs at room temperature. The orthorhombic nature in the Pb-Br-Pb ADF was consistently observed across different supercell sizes including the extrapolative cases (Figure S1). We also assessed vibrational properties by computing the Γ-point phonon density of states (DOS) for a 3×3×3 bare NC as shown in Figure 3c. A linear force-displacement relationship in the harmonic regime was accurately captured by the MLIP, allowing us to compute phonon DOS via the finite difference method (Figure S2). The overall shape of the predicted DOS agrees nicely with the DFT reference, although noticeable peak shifts are present. Similar observations have been reported in other studies,[27,31] underscoring the difficulty of accurately learning the Hessian information, even when force predictions are reliable. In our case, the shifts likely arise from the discrepancies between the bulk and nanocrystal environments that influence the implicitly learned curvature of the potential energy surface. Despite the shifts, the MLIP well reproduces the relative intensities of characteristic low-energy modes, typically associated with $PbBr_6^{4-}$ octahedral rotation (M1) and Pb-Br-Pb stretching (M2 and M3) reported in other computational[10,13,32,33] and experimental[11] studies. A shoulder near 75 cm$^{-1}$ is often attributed to Pb-Br-Pb bending[13]. Interestingly, we find that compared to the DFT peak positions, MLIP peak positions actually align better with the previously reported values[13,32] (M2 ~ 100 cm$^{-1}$ and M3 ~ 129 cm$^{-1}$). These results confirm that the fine-tuned MLIP provides good accuracy in structural and vibrational predictions for both bare and ligand-passivated NCs for further analysis.

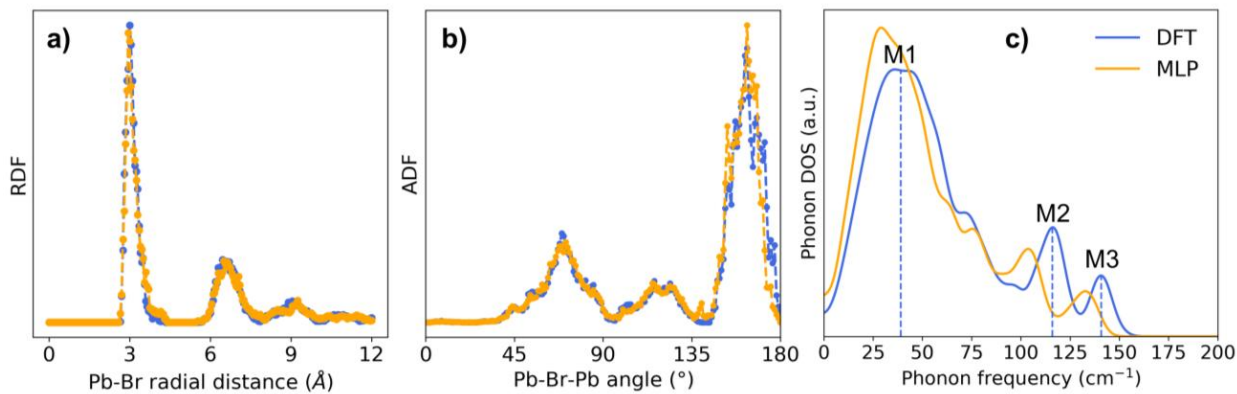

Figure 3. Validation of the fine-tuned MLIP against DFT. (a) Pb-Br radial distribution function, and (b) Pb-Br-Pb angular distribution function, obtained from 2 ps NVT (300K) trajectories of 2×2×2 50% BzO/MA-capped NC. (c) Γ-point phonon DOS of 3×3×3 bare NC, computed via the finite difference method.

**Ligand-controlled phonon dynamics**

We systematically examined the ligand-dependent phonon modes in $CsPbBr_3$ NCs to reveal how surface chemistry influences low-energy lattice dynamics (Figure 4). Phonon DOS was computed for NCs ranging from 3×3×3 (1.79 nm) to 7×7×7 (4.17 nm) with four representative surface terminations: (1) bare, (2) MA-capped, (3) BzO-capped, and (4) BzO/MA-capped. BzO was chosen as a model anionic ligand because of its good passivation capability in $CsPbBr_3$ NCs based on our previous study.[15,34] In all ligand-capped systems, ligand coverage was fixed at 50%, which is close to the upper bound[35–37] of realistic ligand densities in colloidal halide perovskite NCs. The phonon DOS was largely size-independent for both bare[13] and ligand-capped NCs. Across all systems, the Pb/Br partial DOS (dotted) shows that the M2 and M3 modes strictly arise from Pb-Br interactions. On the other hand, the M1 mode is accompanied by the contribution from the Cs atoms, indicating that it involves the collective motions of the entire core atoms coupled with the $PbBr_6^{4-}$ rotations. Upon ligand passivation, the M2 mode went through broadening while M3 stayed relatively well-defined.

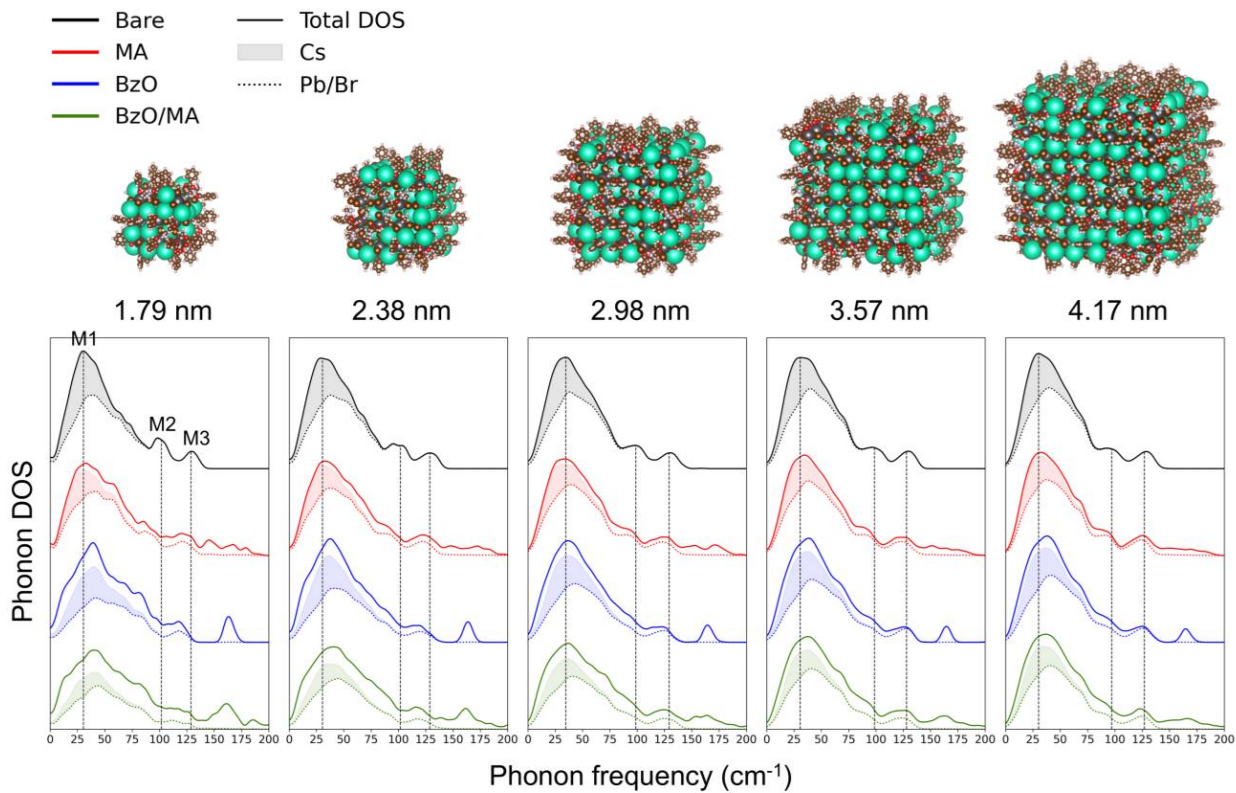

Figure 4. Ligand-dependent phonon DOS in CsPbBr$_3$ NCs. Top: representative structures of MA and benzoate-capped NCs ranging from 3×3×3 to 7×7×7 supercells. Bottom: phonon DOS corresponding to each size for bare (black), methylammonium (MA, red), benzoate (BzO, blue), and BzO/MA (green) passivation. The solid lines represent the total DOS of the system, the dotted lines represent the partial DOS of Pb/Br, and the shaded regions represent the partial DOS of Cs. The black vertical dashed lines indicate low-energy modes (M1, M2, M3) corresponding to PbBr$_6^{4-}$ rotation and Pb-Br-Pb stretching.

When compared to bare surfaces, introducing ligands causes significant redshifts of the Pb-Br-Pb stretching modes (M2 and M3), as evident from the downshift of Pb/Br partial DOS (dotted) in Figure 4. This behavior was consistently observed regardless of the types of surface ligands (i.e., anionic, cationic, or both), but the magnitude of redshift was generally larger for NCs containing anionic ligands (BzO or BzO/MA) compared to NCs only containing cationic ligands (MA), as well depicted in Figure 5. The ligand-induced redshift of Pb-Br-Pb stretching is physically coherent with the nature of chemical bonding. The electron-withdrawing property of anionic ligands pulls away the electron density around the surface Pb atom, reducing the effective Pb-Br bond order. Since the vibrational frequency is proportional to the bond order, the vibrational frequency of the Pb-Br-Pb stretching reduces, which appears as a downshift of M2 and M3. While MA does not directly bond to Pb or Br, its $NH_3^+$ group can form hydrogen bonds with the surface halides,[9,38] as observed in our relaxed structures (Figure S3a). This could have a similar effect in reducing the Pb-Br bond order although its effect would be smaller than that of the anionic ligands. The magnitude of the ligand-induced redshifts tends to diminish with increasing size as the contribution of surface components decreases.

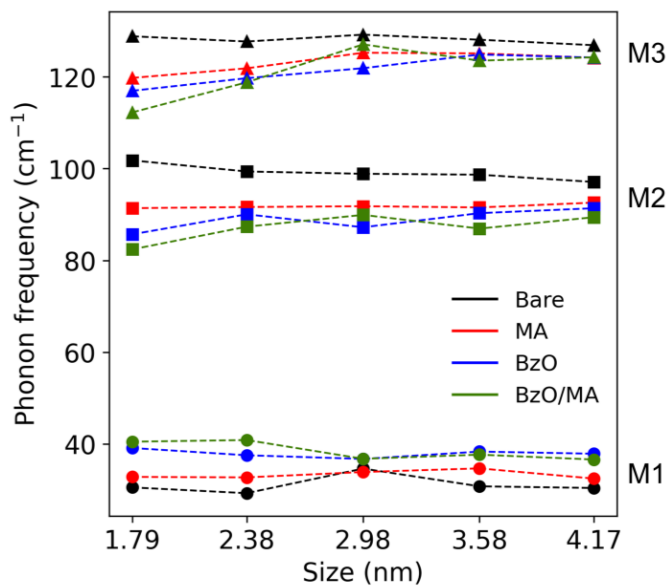

Figure 5. The intensity-weighted peak positions of M1 (circle), M2 (square), and M3 (triangle) across systems with varying sizes and ligand passivations. Black, bare; red, MA; blue, BzO; green, BzO/MA-capped NCs. Ligands induce redshift of the Pb-Br-Pb stretching modes (M2, M3) and blueshift of the $PbBr_6^{4-}$ rotation mode (M1). The effect was generally larger for anionic ligands (blue, green) than cationic ligands (red).

In the BzO/MA-capped systems, the hydrogens in the $-NH_3^+$ group can additionally interact with the BzO (Figure S3b). The hydrogen bonding interaction between BzO and MA is observed as frequency shifts of the pertained ligand modes. In particular, the downshift of the N–H stretching of MA with the presence of BzO (Figure S4) indicates a strong hydrogen bonding interaction between MA and the BzO. The bond order reduction effect of BzO and MA can take place simultaneously in BzO/MA-capped NCs. Hence, the mixed ligand NCs generally have the largest redshifts of M2 and M3. The MA-BzO hydrogen bond also perturbs the ligand-localized mode at ~165 cm$^{-1}$, which consistently appears across all sizes in the BzO-capped NCs. The partial phonon DOS of C, H, O, and N (Figure S5) confirms that this peak strictly pertains to BzO, likely involving out-of-plane bending motion.[39] With the introduction of MA, a significant reduction and broadening of the C and O peaks occur, suggesting a spectral shift of the mode mediated by strong MA-BzO interactions. The role of low-energy ligand modes in optoelectronic properties remains unclear; however, we show that these modes can be effectively stabilized with synergistic surface cation-anion interactions.

Interestingly, the octahedral rotation mode (M1) shows the opposite trend, exhibiting a clear blueshift upon ligand passivation. Consistent with the behavior of M2 and M3, BzO-capped NCs display larger M1 blueshifts than the MA-capped NCs. A similar ligand-induced stiffening of rotational modes was previously reported for FAPbBr$_3$ and attributed to the multipodal hydrogen bonding and Van der Waals interactions between surface FA and bulky ligands.[10] Here, we find that even relatively small ligands such as BzO or MA are sufficient to induce a visible M1 blueshift. In nanocrystals, multiple ligands can coordinate the corner and edge-site Pb atoms, effectively "pinning" them via steric hindrance (Figure S3b). Because M1 corresponds to a collective, low-frequency motion of many atoms, restricting the motion of just a few key surface sites can significantly suppress this global deformation. Likewise, in MA-capped NCs, a dense network of N–H···Br bridge bonds can similarly anchor surface atoms, stiffening the lattice and driving the observed blueshift.

To summarize, our results reveal a clear and mode-selective response of CsPbBr$_3$ nanocrystals to surface passivation. Local Pb-Br-Pb stretching modes (M2 and M3) are softened through ligand-induced reduction of Pb-Br bond orders, whereas the collective octahedral rotation mode (M1) is stiffened through steric pinning and hydrogen-bond-mediated anchoring at surface lattice sites. The contrasting redshift-blueshift behavior highlights the dual role of ligands: they electronically weaken local metal-halide bonding while simultaneously imposing structural constraints that suppress low-energy lattice fluctuations. This interplay between local chemical bonding and global lattice rigidity illustrates how surface chemistry can be leveraged to selectively control lattice dynamics of halide perovskite nanocrystals.

**The role of ligand binding energies in M1 blueshift**

The blueshift of the M1 mode suggests the suppression of global dynamic disorder, which is associated with detrimental nonradiative recombination in halide perovskite NCs. The ligand-induced M1 blueshift has already been shown beneficial to improve the PLQY significantly.[10] Based on our observations that anionic ligands induce a larger shift, we further investigated the effect of different anionic ligands on realistic nanocrystal surfaces. In addition to benzoate (BzO), we considered thiophenolate (PhS) and phenylphosphonate (PhP), which share a benzene backbone but differ in their binding motifs and binding energies to surface Pb atoms (Figure S6). Figure 6 shows the intensity-weighted peak positions of the M1 mode in PhS/MA, BzO/MA, and PhP/MA-capped NCs. The introduction of anionic ligands induced blueshifts of the M1 mode compared to the bare surface baseline (black). Interestingly, we observed that the magnitude of the blueshift did not follow a monotonic trend with the ligand binding energy. In fact, BzO, which has the binding energy closest to the native Br atom, consistently induced the most blueshift across all sizes, compared to the stronger-binding PhP or weaker-binding PhS. The same trend was also observed in the NCs passivated only with anionic ligands (Figure S7), suggesting a systematic effect of the anionic ligands on the M1 blueshift. Given that the size and the steric pinning effect of the three anionic ligands are comparable, we conclude that the blueshift of the $PbBr_6^{4-}$ rotation mode is sensitive to the chemical properties of the anionic ligands coordinating the Pb atoms.

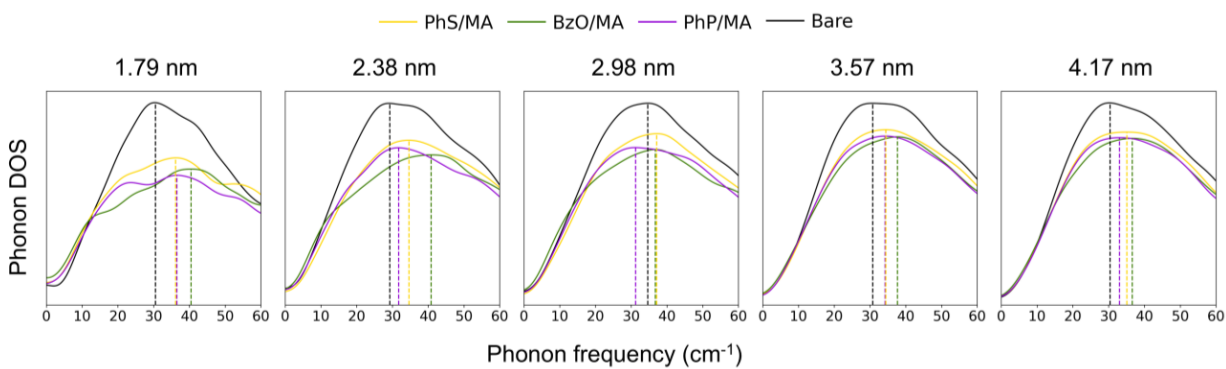

Figure 6. The M1 mode in $CsPbBr_3$ NCs with bare surface (black) and 50% coverage PhS/MA (yellow), BzO/MA (green), and PhP/MA (purple). The vertical dotted lines show the intensity-weighted peak position of the M1 mode. The ligand-capped NCs show consistent blueshift across all sizes compared to the bare NCs. BzO induces the largest blueshift systematically throughout all systems studied.

To gain further mechanistic insights into this non-monotonic trend, we performed room-temperature molecular dynamic simulations for small 3×3×3 mixed-ligand NCs, where most atoms reside on the surface. We then computed the root mean squared fluctuations (RMSF) of the Pb atoms, which sit at the center of the $PbBr_6^{4-}$ octahedra (Figure 7). Notably, the BzO/MA system exhibited both the lowest average RMSF and the narrowest fluctuation distribution, indicating that the Pb atoms are more spatially constrained. In contrast, PhS/MA and PhP/MA-capped NCs showed larger average deviations and broader error bars, suggesting greater dynamical disorder within the inorganic framework. This observation aligns with our phonon calculations and can be rationalized by a chemical-bond conservation principle: when the ligand-Pb interaction is either too weak or too strong relative to the intrinsic Pb-Br bonding, it disrupts lattice equilibrium and enhances atomic motion. Weak binding leaves surface Pb sites insufficiently stabilized, whereas overly strong binding over-constrains or distorts neighboring Pb-Br bonds, both leading to excessive bond-length fluctuations. Ligands with intermediate binding strength, comparable to that of the native halides, best preserve this balance, maintaining uniform Pb-Br interactions and minimizing lattice disorder. Consequently, such ligands can suppress low-energy lattice fluctuations that serve as nonradiative recombination channels for charge carriers.

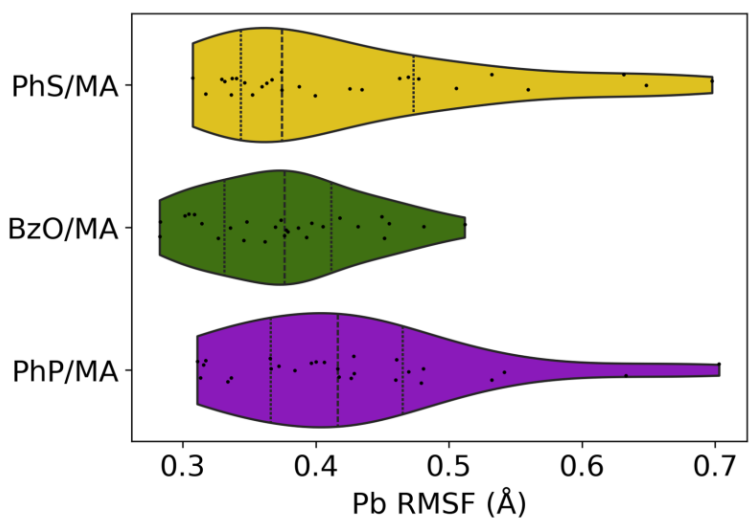

Figure 7. The root mean squared fluctuations (RMSF) of the Pb atoms in the representative 3×3×3 systems, obtained from the 40 ps NVT trajectories at 300K.

## CONCLUSIONS

In conclusion, we have shown that surface ligand plays a critical role in modulating low-energy phonons of $CsPbBr_3$ nanocrystals. By fine-tuning a universal MLIP on small, ligand-passivated

NCs, we systematically explored realistic NC sizes and diverse ligand chemistries, which were previously infeasible with conventional *ab initio* methods. We uncover a mode-selective response in which local Pb-Br-Pb stretching (M2, M3) softens due to ligand-induced reduction of Pb-Br bond order, while the collective octahedral rotation (M1) stiffens via steric pinning and hydrogen-bond anchoring. The cationic and anionic ligands can have synergistic effects on the surface by forming hydrogen bonds, which help stabilize ligand-localized low-energy modes and strengthen lattice stiffening. The M1 blueshift exhibits a non-monotonic dependence on anionic ligand binding energy: benzoate, which has the closest binding energy to native halides, most effectively damps rotational disorder, whereas much stronger (PhP) or weaker (PhS) binding ligands could be suboptimal, as corroborated by molecular dynamics. Our work provides insights into ligand-controlled phonon dynamics in realistic halide perovskite nanocrystals, which have important implications for the radiative performance of these materials. More broadly, our work establishes fine-tuned MLIP as a powerful tool to bridge theoretical insights and practical design rules for next-generation perovskite nanocrystal optoelectronics.

## ACKNOWLEDGEMENTS

This material is based upon work supported by the U.S. National Science Foundation under Grant Number 2339866, and the U.S. National Science Foundation Science and Technology Center (STC) for Integration of Modern Optoelectronic Materials on Demand (IMOD), under Cooperative Agreement No. DMR-2019444. This research used resources of the National Energy Research Scientific Computing Center (NERSC), a Department of Energy User Facility using NERSC award BES-ERCAP35988. DFT research conducted as part of a user project at the Center for

Online Supporting Information for

# Ligand-Controlled Phonon Dynamics in CsPbBr$_3$ Nanocrystals Revealed by Machine-Learned Interatomic Potentials

Seungjun Cha,[1] Chen Wang,[2,3,*] Victor Fung,[4,*] Guoxiang Hu[1,5,*]

[1] School of Materials Science and Engineering, Georgia Institute of Technology, Atlanta, GA 30332, USA

[2] Department of Chemistry and Biochemistry, Queens College, City University of New York, New York, NY 11367, USA

[3] The Graduate Center, City University of New York, New York, NY 10016, USA

[4] School of Computational Science and Engineering, Georgia Institute of Technology, Atlanta, GA 30332, USA

[5] School of Chemistry and Biochemistry, Georgia Institute of Technology, Atlanta, GA 30332, USA

*Corresponding author.

Email: chen.wang@qc.cuny.edu, victorfung@gatech.edu, emma.hu@mse.gatech.edu

# METHODS

## 1. Model generation

Cubic CsPbBr$_3$ nanocrystals (NCs) with dimensions of $n \times n \times n$ unit cells ($n$ = 3, 4, 5) were constructed with a lattice parameter of 5.95 Å using Pymatgen[1] (v2024.11.13). All surfaces were initially terminated with Cs and Br atoms[2,3]. The net charge of each NC was obtained by summing over the formal ionic charges of Cs$^+$, Pb$^{2+}$, and Br$^-$. For charge neutrality, Cs atoms were first removed from all corner sites and then randomly deleted from remaining sites as needed. Organic ligands were generated from their SMILES representations and geometry-optimized with the Merck Molecular Force Field (MMFF)[4] as implemented in RDKit[5] (v2024.03.5). Each ligand was oriented such that its binding motif was perpendicular to a {100} surface, while retaining rotational freedom about the surface normal. Cationic ligands were used to replace surface Cs sites, and anionic ligands replaced surface Br sites, to achieve specified coverages ranging from 20% to 60%. For mixed-ligand NCs, cationic–anionic ligand pairs were placed adjacently[6]. After ligand placement, a large vacuum of 15 Å was introduced along all three Cartesian directions to avoid interactions between periodic images. A random perturbation proportional to the covalent radius of each atom was applied: 10–15% for C, H, O, N, S, and P, and 15–25% for Cs, Pb, and Br. For mixed-ligand NCs, a short pre-relaxation was performed using the pretrained MatterSim-v1.0.0-1M[7] until the maximum force was below 0.5 eV/Å to prevent steric crowding prior to density functional theory (DFT) geometry relaxation. In total, 236 NC models were generated with varied supercell sizes, ligand coverages, and surface passivation schemes.

## 2. DFT geometry relaxation

Density Functional Theory (DFT) calculations were performed using the Vienna *Ab initio* Simulations Package[8] (VASP, v.6.4.2). The ion-electron interaction was described by the projector-augmented wave (PAW) method[9], with a planewave cutoff energy set to 400 eV. The exchange-correlation interaction was described by the generalized gradient approximation (GGA) method of Perdew, Burke, and Ernzerhof (PBE)[10]. All nanocrystal (NC) models were treated non-periodically with a large vacuum of 15 Å in all directions, so a Γ-point (1 × 1 × 1) Monkhorst–Pack *k*-point mesh was used. Spin–orbit coupling (SOC) was ignored in these calculations. The D3 dispersion energy-corrections were not included at this stage but in the later steps *post hoc* (See Section 4). The energy convergence criterion of $1\times10^{-4}$ eV and the force convergence criterion of 0.03 eV/Å were used. The precision and electronic minimization algorithms were set to PREC = Normal and ALGO = Fast, respectively. A maximum of 60 electronic steps per ionic step (NELM = 60) was allowed. Due to the large computational cost of relaxing NC models, each structure was relaxed either until full convergence or until a wall-time limit of 6 hours was reached, whichever is earlier. As a result, some NCs were only partially relaxed during the dataset generation. All validation cases were converged fully. On average, each run went through ~240 ionic steps. All calculations were performed using 8 NVIDIA A100 GPUs per run.

## 3. Fine-tuning machine learning potential

56,647 structural snapshots that electronically converged (< NELM) were extracted from the DFT geometry relaxation trajectories. The dataset was randomly split into training (68%), validation (12%), and test (20%) sets. Only total energies and atomic forces were included. All layers of the MatterSim-v1.0.0-1M checkpoint were fine-tuned using the MatterTune[11] (v.0.1.0) package. The two-body cutoff and three-body cutoffs were set as 6.5 Å and 4.0 Å, respectively, to sufficiently capture the Pb–Br bonds and Pb–Br–Pb angles within the radial cutoffs. The remaining graph conversion hyperparameters were set to the MatterSim default. The loss function was the mean squared error (MSE) of energies and forces with the relative weight of 1 and 10, respectively. The model was optimized using AdamW[12] with the initial learning rate $1\times10^{-4}$ and the cosine annealing as a scheduler (final learning rate of $1\times10^{-6}$). Training was done for 350 epochs with batch size 8 and 16-mixed precision on 4 NVIDIA V100 GPUs. The best model was selected based on the lowest validation force MAE.

## 4. Molecular dynamics

*Ab initio* molecular dynamics simulations were performed in the canonical ensemble (NVT) using the Langevin thermostat as implemented in VASP. The target temperature was set to 300 K, and initial velocities were sampled from a Maxwell–Boltzmann distribution. A timestep of 1 fs was chosen in accordance with the highest vibrational modes of hydrogen. The Langevin damping constant was set to 20 ps$^{-1}$ for all elements (LANGEVIN_GAMMA = 20) for quick thermal equilibration. The energy convergence criterion was tightened to $1 \times 10^{-6}$ eV, and precision was set to PREC = Accurate for numerical stability. All other settings were consistent with the geometry relaxations. The duration was typically set to ~2 ps (2000 steps), which was sufficient to sample near-equilibrium phase space.

Molecular dynamics with fine-tuned MLP was performed using the Atomic Simulation Environment[13] (ASE, v3.24.0) interface provided in MatterTune. For validation studies, all simulation settings were the same as those used for *ab initio* simulations. For long MD (~40 ps), D3 dispersion-energy corrections[14] with Becke and Johnson damping[15] were additively applied to stabilize energy drifts using the ASE SumCalculator interface implemented in Simple-DFTD3 package[16]. Snapshots were saved every 50 fs and those in the first 10 ps were discarded for thermal equilibration. Only the remaining snapshots were used to compute the radial distribution function (RDF) and the angular distribution function (ADF) using Auto-FOX[17].

## 5. Phonon calculations

For the DFT workflow, we started from the DFT-relaxed geometries without further tightening the force convergence criteria, as stricter relaxation yielded only marginal differences in the phonon DOS while significantly increasing the computational cost (see Figure S8). Here, Phonopy[18] (v.2.38.0) was used to create displaced structures and the force constant matrix. Each atom in the relaxed structure was displaced by $\pm 0.01$ Å in the $x, y, z$ directions, creating $6N$ displaced structures, where $N$ is the total number of atoms. For each displaced structure, a single-point DFT calculation was performed in VASP using the same settings as the geometry relaxations but with the tighter energy convergence criterion of $1 \times 10^{-8}$ eV, PREC = Accurate, and ALGO = Normal for improved force precision. The force constant matrix was computed by the finite difference method as implemented in Phonopy, which was then passed to ASE's Phonons module to compute the Γ-point phonon DOS for easier workflow integration. The partial DOS was computed by projecting eigenvector amplitudes of the specified atoms onto the total DOS then normalizing by atomic mass. The phonon DOS was smoothened by assigning each eigenfrequency a normalized Gaussian ($\sigma = 3.0$ cm$^{-1}$) and summing over all modes. Peak positions were obtained from the intensity-weighted center, using a frequency window defined by curvature changes in the spectrum. For M1 mode, a fixed $\pm 10$ cm$^{-1}$ frequency window around the maximum peak was used.

The ML-based phonon workflow was carried out entirely within ASE. In this case, the starting geometry was either taken directly after DFT geometry relaxation (for validation cases) or relaxed using the L-BFGS algorithm[20] with the fine-tuned MLP. The subsequent displacement, force constant computation, and DOS calculation followed the same procedures using the ASE Phonons module. The D3 dispersion was ignored during the phonon workflow because the force contribution of the D3 dispersion at finite displacements was orders of magnitude smaller than the MLP forces.

## Supplementary Figures and Tables

**Table S1** A full list of anionic ligands used for modelling mixed-ligand NCs.

| Ligands | Molecular structure |
| --- | --- |
| Octylphosphonate (OP) | 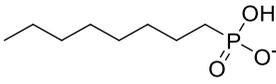 |
| Phenylphosphonate (PhP) | 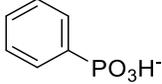 |
| 2-Pyridinesulfonate (PyS) | 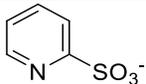 |
| Dodecyl sulfate (DS) | 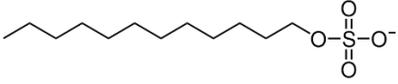 |
| Dodecyl benzenesulfonate (DBS) | 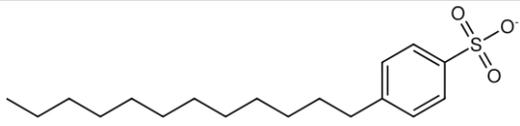 |
| Benzenesulfonate (BS) | 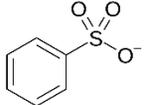 |
| Pyrrole-2-carboxylate (Pyr-2-COO) | 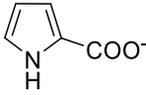 |
| 2-Furoate (2-Fur) | 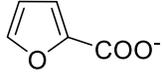 |
| 2-Thiophenecarboxylate (TPC) | 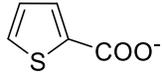 |
| 4-Bromothiophene-2-carboxylate (4-Br-2-TPC) | 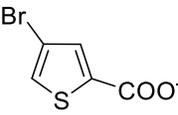 |
| Benzoate (BzO) | 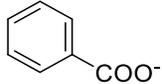 |
| Isonicotinate (INC) | 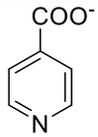 |

| Name | Structure |
|---|---|
| Anthranilate (ANT) | 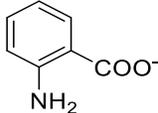 |
| Picolinate (PIC) | 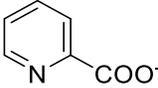 |
| 5-chloropyridine-2-carboxylate (5-Cl-PIC) | 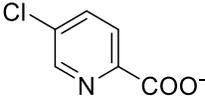 |
| Fusarate (FUS) | 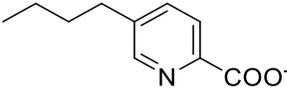 |
| Pyridine (Py) | 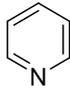 |
| 2-Mercaptopyridine (2-MPy) | 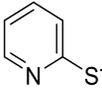 |
| Thiophenolate (PhS) | 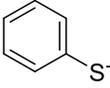 |
| Thiophene (Th) | 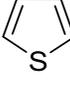 |
| Anthranilic acid (2-ABA) | 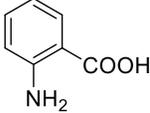 |
| 2,5-pyrazinedicarboxylic acid (PDC) | 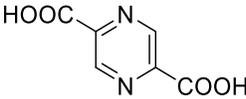 |
| L-cysteine (Cys) | 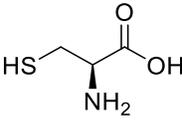 |
| Glutathione (GSH) | 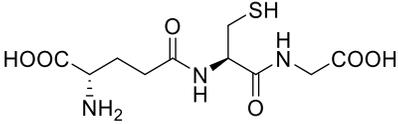 |

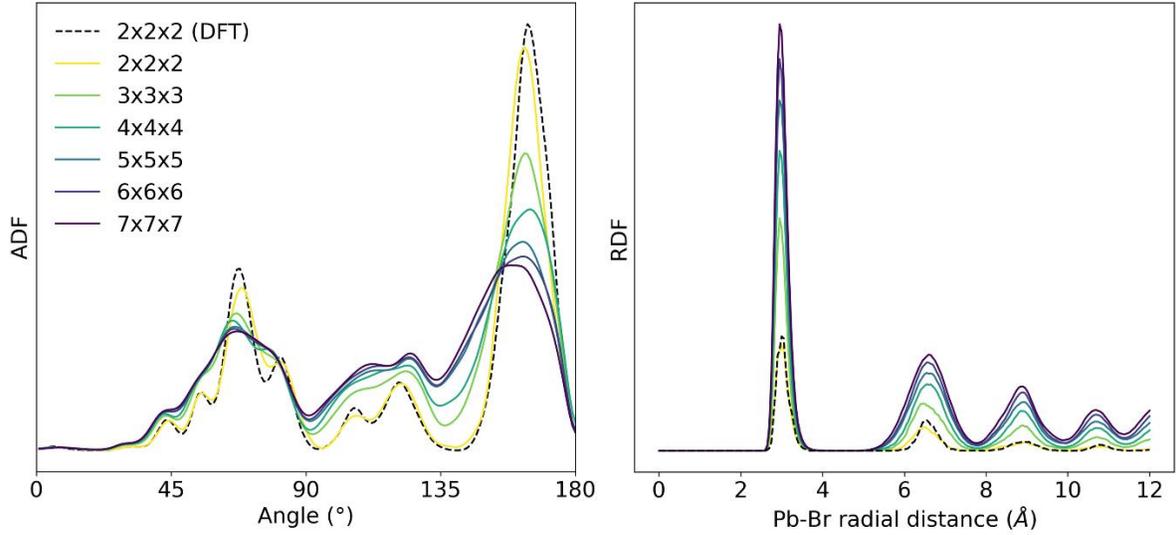

**Figure S1.** Angular distribution functions of bare CsPbBr$_3$ with varying supercell sizes (2×2×2 to 7×7×7), obtained from 10 ps NVT trajectories at 300K. The fine-tuned MLP well reproduced the orthorhombic nature of CsPbBr$_3$ at room temperature across all sizes.

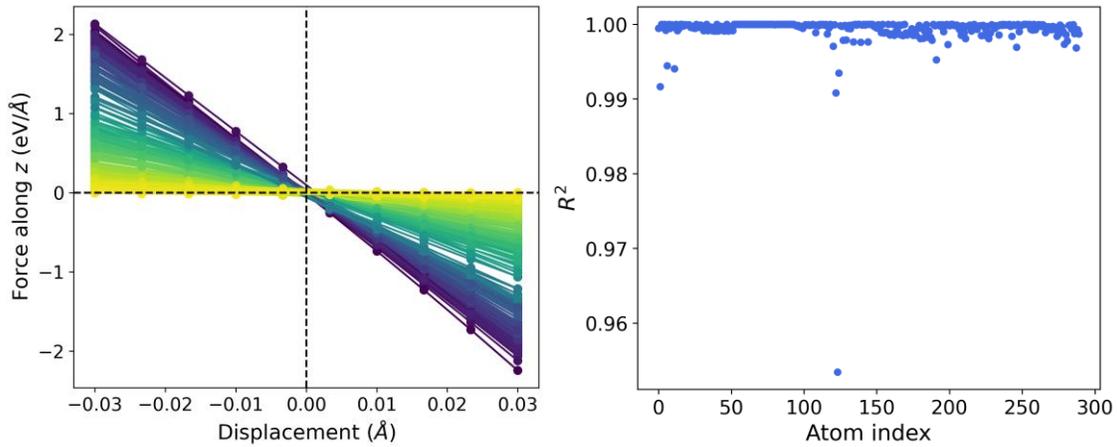

**Figure S2.** Harmonic force–displacement response computed from the fine-tuned MLP for a 2×2×2 CsPbBr$_3$ capped with 50% BzO/MA. Each colored line (left) corresponds to the force response of a single atom under small displacements along the $z$ axis. The scatter plot (right) shows the $R^2$ of the linear fit for each line. The colormap is proportional to the slope of the fitted line and is shown only for visualization. The linear relation demonstrates that the fine-tuned MLP can accurately capture the harmonic vibrational behavior, which is required for phonon DOS calculations via the finite difference method.

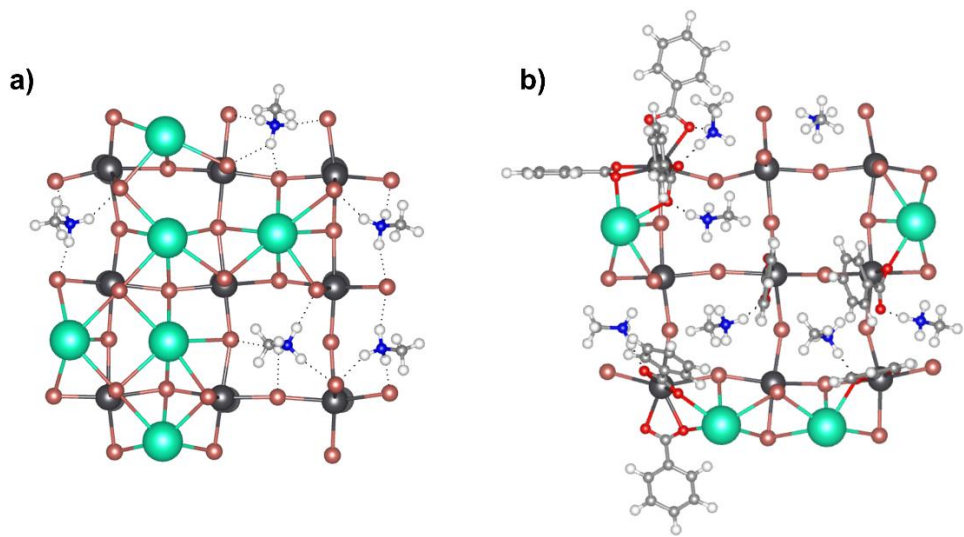

**Figure S3.** Optimized atomic structures of (a) MA-capped, and (b) BzO/MA-capped NC surfaces. H, white; C, light grey; N, blue; O, red; Br, brown; Pb, dark grey; Cs, green. Dotted lines represent hydrogen bonds between H–Br (a), or H–O (b).

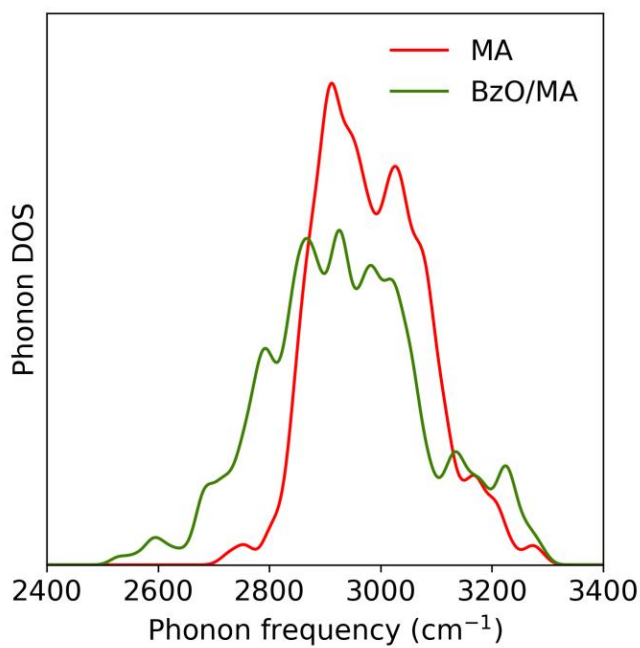

**Figure S4.** The N–H stretching mode in the 7×7×7 MA and BzO/MA-capped NCs. A clear downshift and broadening of the N–H stretching in the mixed-ligand NCs suggest a strong hydrogen bond between the cationic and anionic species.

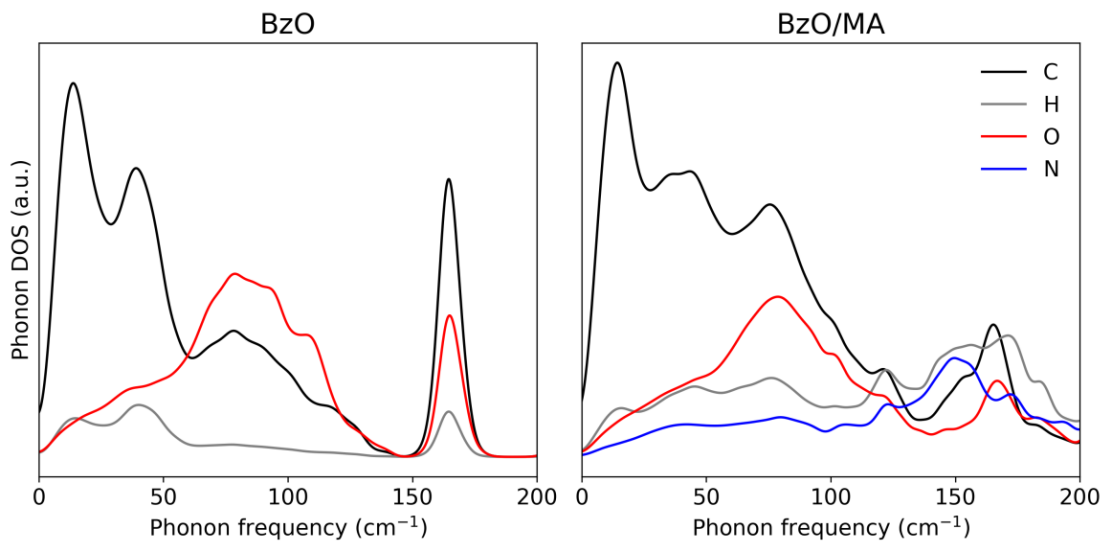

**Figure S5.** Partial DOS of C, H, O, and N for 7×7×7 MA and BzO/MA-capped NCs. A significant reduction and broadening of the C and O peaks are observed in the mixed-ligand NCs, suggesting a spectral shift of the mode mediated by strong MA–BzO interactions.

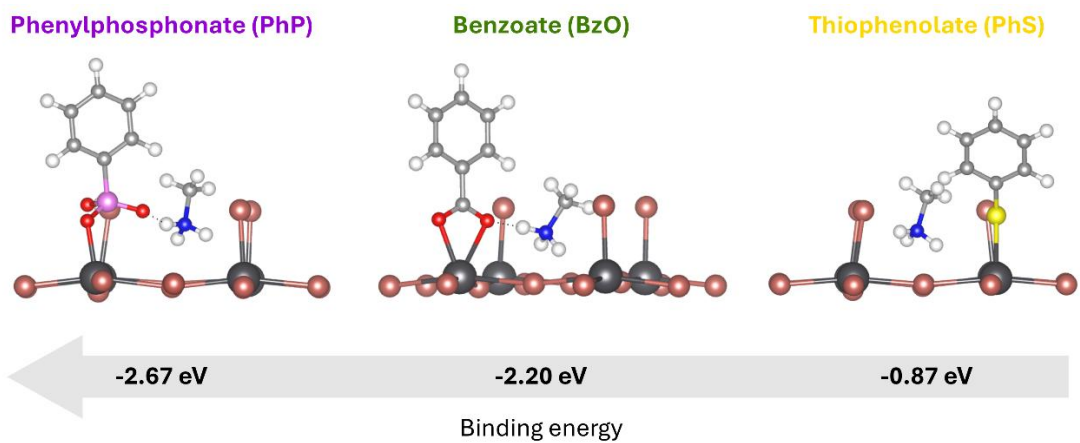

**Figure S6.** Optimized atomic structures of a CsPbBr$_3$ surface capped with PhP/MA (left), BzO/MA (middle), and PhS/MA (right). H, white; C, light grey; N, blue; O, red; P, pink; S, yellow; Br, brown; Pb, dark grey. The dotted line represents hydrogen bonds. The DFT-calculated binding energy corresponds to that of each anionic ligand with respect to the 2×2×3 MA-capped CsPbBr$_3$ slab, where the binding energy of Br$^-$ is -2.01 eV.

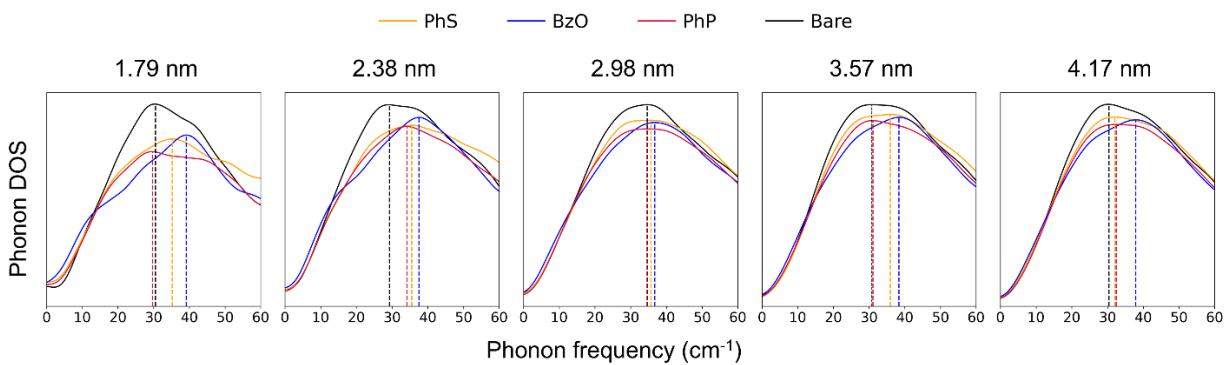

**Figure S7.** Ligand-dependent M1 mode in CsPbBr$_3$ NCs with 50% coverage PhS (orange), BzO (blue), PhP (brown). The vertical dotted lines show the intensity-weighted peak position of the M1 mode.

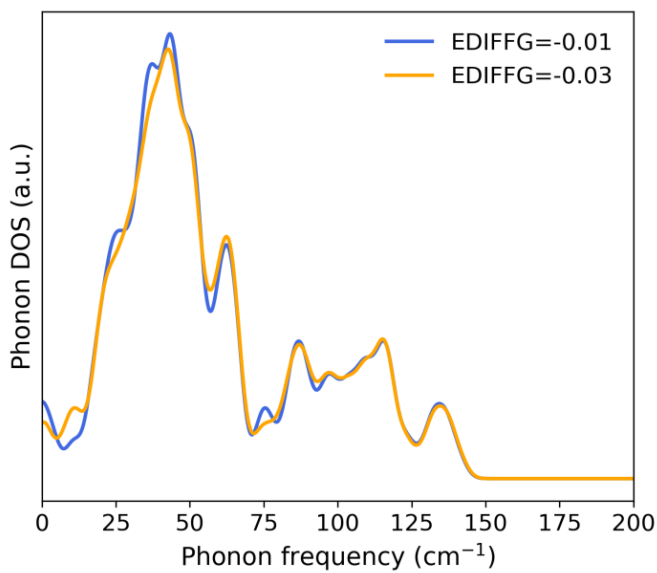

**Figure S8.** The *ab initio* phonon DOS of 2×2×2 bare CsPbBr$_3$ NC with the force criterion (EDIFFG) of -0.01 and -0.03. A stricter convergence criterion beyond -0.03 made negligible differences while adding significant computational cost for relaxing free-floating nanocrystals.